\documentclass[twoside,twocolumn]{article}

\usepackage[sc]{mathpazo} 
\usepackage[T1]{fontenc} 
\usepackage{lmodern} 
\usepackage{microtype} 

\usepackage[english]{babel} 

\usepackage{graphicx}
\usepackage{color}
\usepackage[top=2cm,bottom=2cm,left=1.5cm,right=1.5cm,columnsep=20pt]{geometry} 
    {\large\scshape Computational Details%
    \par\medskip\normalfont\normalsize}%
    {}%
\newenvironment{acknowledgement}
    {\large\scshape Acknowledgement%
    \par\medskip\normalfont\normalsize}%
    {}%
    {\large\scshape Supporting Information%
    \par\medskip\normalfont\normalsize}%
    {}%
\usepackage{multirow} 
\usepackage{slashbox} 

\usepackage[normal, small,labelfont=bf,up,textfont=it,up]{caption} 
\usepackage{booktabs} 

\usepackage{lettrine} 

\usepackage{enumitem} 
\setlist[itemize]{noitemsep} 

\usepackage{abstract} 

\usepackage{titlesec} 
\renewcommand\thesection{\Roman{section}} 
\renewcommand\thesubsection{\roman{subsection}} 
\titleformat{\section}[block]{\large\scshape\centering}{\thesection.}{1em}{} 
\titleformat{\subsection}[block]{\large}{\thesubsection.}{1em}{} 

\usepackage{fancyhdr} 
\usepackage{titling} 

\usepackage{hyperref} 

\pretitle{\begin{center}\Large\bfseries} 
\posttitle{\end{center}} 

\usepackage{amsmath}
\usepackage{doi,hyperref}
\usepackage[numbers]{natbib}
\usepackage{chemformula}
\usepackage{ulem}
\usepackage{xr}
\newcommand{\beginsupplement}{%
        \setcounter{table}{0}
        \renewcommand{\thetable}{S\arabic{table}}%
        \setcounter{figure}{0}
        \renewcommand{\thefigure}{S\arabic{figure}}%
     }

\def\sss{\scriptscriptstyle\rm}

\def\b{_{\sss b}}

\def\LDA{^{\sss LDA}}
\def\HF{^{\sss HF}}

\def\33{sp^3$-$sp^3}
\def\mix{sp^3$-$sp^2}

\def\22{sp^2$-$sp^2}
\def\ot{_{\sss ot}}

\title{Explaining and Fixing DFT Failures for Torsional Barriers} 
\author{%
\textsc{Seungsoo Nam, Eunbyol Cho, and Eunji Sim}\thanks{esim@yonsei.ac.kr} \\ 
\normalsize Department of Chemistry, Yonsei University, 50 Yonsei-ro Seodaemun-gu, Seoul 03722, Korea \\ 
\textsc{Kieron Burke} \\ 
\normalsize Departments of Chemistry and Physics, University of California, Irvine, CA 92697, USA \\  
}
\date{\today} 

\renewcommand{\maketitlehookd}{
\begin{abstract}
\noindent 
Most torsional barriers are predicted to high accuracy (about 1~kJ/mol) by standard semilocal functionals, but a small subset has been found to have much larger errors. We create a database of almost 300 carbon-carbon torsional barriers, including 12 poorly behaved barriers, all stemming from Y=C-X group, where X is O or S, and Y is a halide. Functionals with enhanced exchange mixing (about 50\%) work well for all barriers. We find that poor actors have delocalization errors caused by hyperconjugation. These problematic calculations are density sensitive (i.e., DFT predictions change noticeably with the density), and using HF densities (HF-DFT) fixes these issues. For example, conventional B3LYP performs as accurately as exchange-enhanced functionals if the HF density is used. For long-chain conjugated molecules, HF-DFT can be much better than exchange-enhanced functionals. We suggest that HF-PBE0 has the best overall performance.
\end{abstract}
}

\begin{document}

\maketitle

\sf

The accurate prediction of the torsional energy landscape of a molecule plays a crucial role in chemical and biological processes, such as estimating the selectivity of chemical reactions,\cite{NBW02,CYDH06,WKOH12,MBC18,WH14} protein folding,\cite{LPPM10,RTJ15}
drug design through docking simulations,\cite{RWDR19,WCLB19} and molecular electronics,\cite{ZB07,VMEN09,CKKK12,LHN18,LJLD17,LSP19} etc.
Density functional theory (DFT) calculations play an essential role in estimating the torsional energy profile\cite{NBW02,VMEN09,CKKK12} and also serve as a benchmark for force-field parameterization in molecular dynamics simulations.\cite{PB10,RTJ15,RWDR19}
Standard DFT calculations using semilocal functionals and hybrids such as B3LYP,\cite{B3LYP} achieve useful accuracies for the torsional profile, showing an error of less than 2~kJ/mol for typical torsional barriers.\cite{CKK97,KCK97,GMTKN55}

However, DFT is known to inaccurately estimate some torsional energies quantitatively and sometimes even qualitatively, particularly for $\pi$-conjugated molecules.\cite{CKK97,KCK97,SPM01,ZT06, SCB10, SKGB14, BW14,TBRL18}
The delocalization error in DFT that overstabilizes the delocalized electronic structure\cite{CMY08} is known to be the cause of these inaccuracies.\cite{PRCS09}
This delocalization error occurs because the delocalized exact exchange hole in the $\pi$-conjugated system is poorly described by local- or semi-local exchange hole models.\cite{SPM01}
By increasing the exact exchange portion or tuning the range-separation parameters in range-separated DFT,\cite{ITYH01} this problem can be partially solved.\cite{SKGB14,KB14,LJLD17}
Tahchieva et~al. could also fix poor barriers from standard DFT with empirical corrections.
These quantitative errors sometimes yield qualitatively incorrect results.
For example, standard DFT overpredicts the torsional profile of resorcinol, producing incorrect trajectories in molecular dynamics simulations.\cite{BVLM20}
Other studies showed that relative energies between two torsional conformations of methyl vinyl ketone cation predicted using popular B3LYP\cite{B3LYP} or CAM-B3LYP\cite{CAMB3LYP} functionals do not match experiment.\cite{PCKK20}

\begin{figure}[h!]
\centering
\includegraphics[width=\columnwidth]{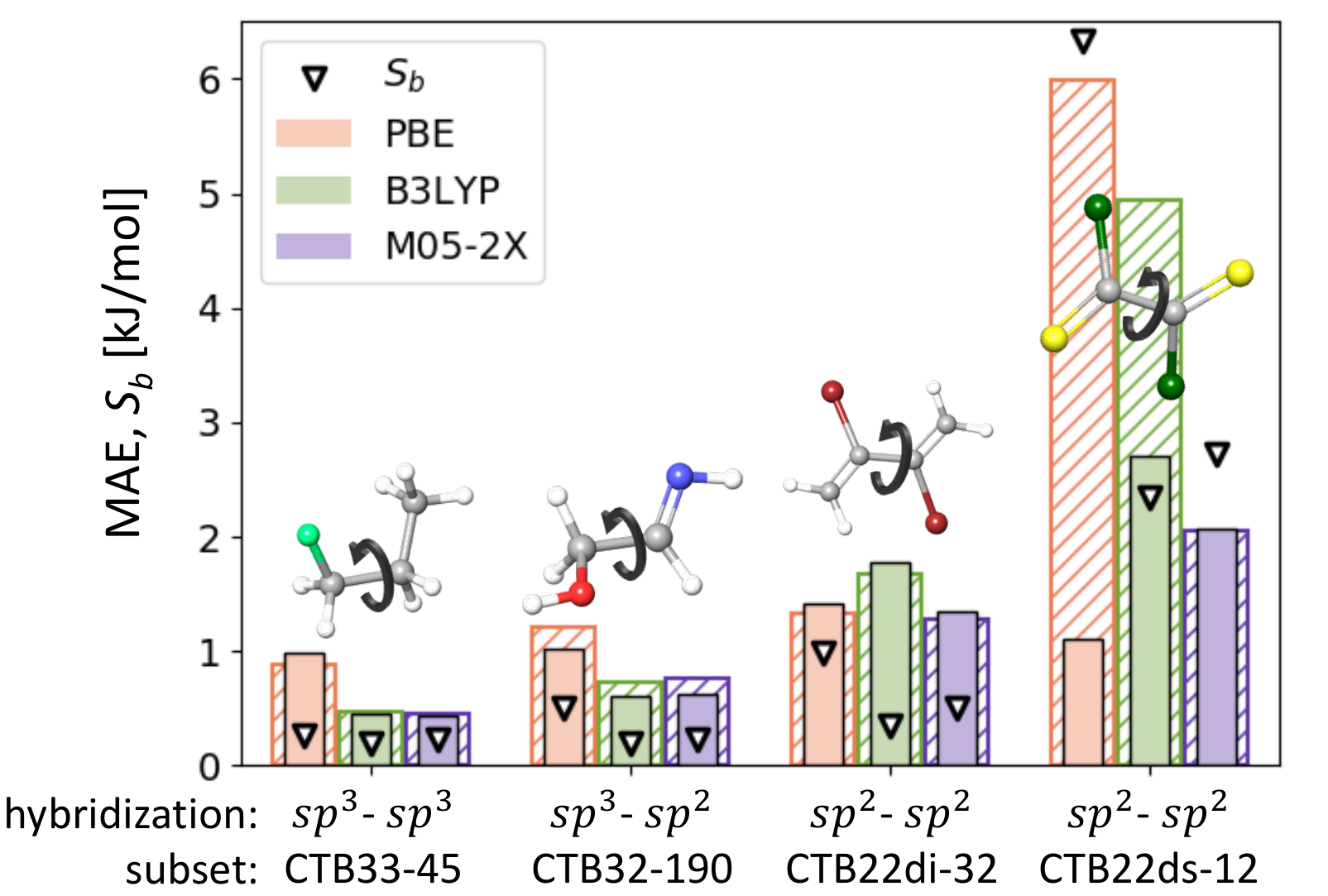}
\caption{
Barrier mean absolute error (MAE) and barrier sensitivity ($S\b$) of subsets in CTB-279.
(Hatched bars denote self-consistent DFT, filled bars HF-DFT.)
CTB$xy$-$N$ consists of $N$ torsional barriers with an axis of torsional rotation between the carbon ($sp^x$)-carbon($sp^y$) bond.
CTB22ds-12 is a subset of CTB22-44 composed of barriers with high barrier sensitivity, and CTB22di-32 contains the rest.
The triangle denotes average barrier sensitivity $S\b$ for each functional.
Representative molecules and their single bonds of interest are also depicted.
}
\label{fgr:summary}
\end{figure}

In this study, we explain the source of the DFT torsional barrier error and propose a simple method to improve various carbon--carbon torsional barriers, including long polymers.
First, we create a carbon torsional barrier (CTB) dataset consisting of 279 torsional barriers with various functional groups attached to carbon atoms.
(See methods for details)
This CTB-279 dataset is further classified into CTB$xy$-$N$ ($x, y=2, 3$, and $N$ represents the number of barriers in a subset) subsets based on the hybridization of the carbon atoms ($sp^2$ or $sp^3$) that make up the torsional axis of rotation. 
For example, CTB22-44 consists of 44 torsional barriers of short, conjugated molecules with a torsional rotation of C($sp^2$)-C($sp^2$).
We further divided the CTB22-44 subset into CTB22di-32 and CTB22ds-12, where CTB22ds-12 is composed of barriers with high barrier sensitivity, and CTB22di-32 is the rest.
(The definition of barrier sensitivity and criteria for this division will be explained later.)
The hatched bars in Figure~\ref{fgr:summary} indicate the mean absolute errors (MAEs) of various levels of functional approximations.
Regardless of the level of approximation, the MAE mildly increases with the number of double bonds (i.e., $\33 \rightarrow \mix \rightarrow \22$).
But the DFT performance is significantly different in the CTB22di-32 and CTB22ds-12 sub-subsets. 
In CTB22ds-12 sub-subset, the generalized gradient approximation and global hybrids perform particularly poorly, whereas M05-2X,\cite{M052X} with a high portion of exact exchange, and the long-range corrected $\omega$B97 series perform well. (See Figure~\ref{S-fgr:stats} and Table~\ref{S-table:MAE} for statistics.)
$\rightarrow$ We also demonstrate a simple remedy for reducing DFT errors: the use of HF-DFT,\cite{HF-DFT}
The solid bars in Figure~\ref{fgr:summary} show that the HF-DFT versions of PBE\cite{PBE96} and B3YLP\cite{B3LYP} (namely, HF-PBE and HF-B3LYP, respectively) now perform similarly to M05-2X (on which the application of HF-DFT has almost no effect).\cite{KSSB18}
Because M05-2X is recommended for a specific purpose,\cite{VT20} HF-DFT is a more general alternative that can be applied to systems that include not only torsional barriers but also the type of interactions for which M05-2X may fail.\cite{PS12}
The rest of this paper explains these results and shows their implications.

\begin{figure}[h!]
\centering
\includegraphics[width=\columnwidth]{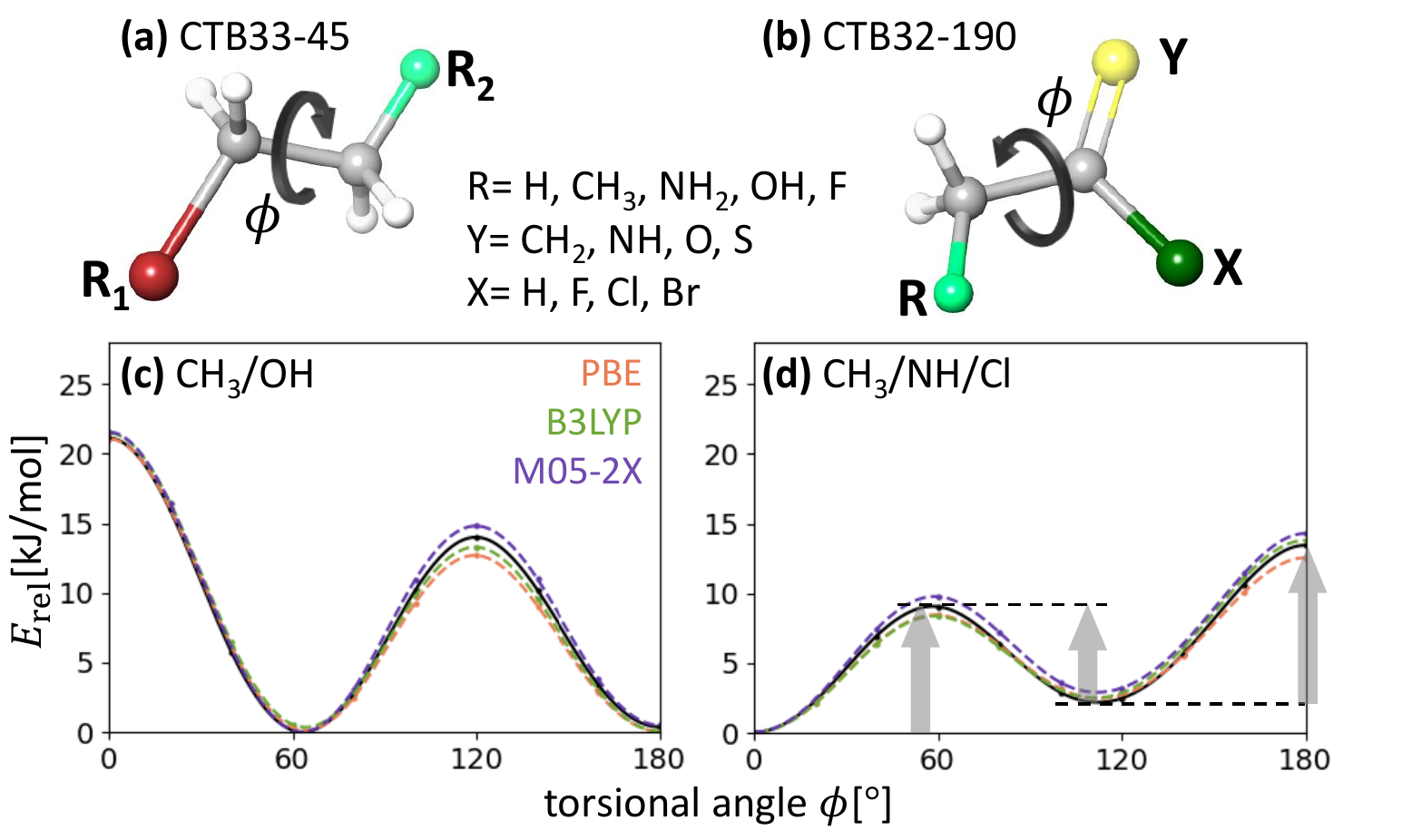}
\caption{
Quality of typical semilocal functionals for typical torsional barriers.
Schematics of molecules in the (a) CTB33-45 and (b) CTB32-190 subsets.
Torsional angle $\phi$ is defined as the R$_1$-C-C-R$_2$ dihedral angle for (a) and R-C-C-Y dihedral angle for (b).
Typical torsional profiles of molecules in the (c) CTB33-45 and (d) CTB32-190 subsets, with reference marked in black.
For each profile, barriers are defined as the energy difference between the local minimum and nearest local maximum, depicted with a gray arrow.
}
\label{fgr:noconj}
\end{figure}

As a torsional axis of rotation, molecules in the CTB-279 dataset contain two single-bonded carbon atoms.
To cover diverse chemical compositions around the carbon atoms, various functional groups are attached.
Figures~\ref{fgr:noconj}a and b show the schematics of molecules in the CTB33-45 and CTB32-190 subset, respectively, and Figures~\ref{fgr:noconj}c and d depict the corresponding torsional profiles.
Torsional barriers are defined as the energy differences between the local minimum and nearest local maximum of each torsional profile, as depicted in Figure~\ref{fgr:noconj}d.

As shown in Figures~\ref{fgr:noconj}c and d, DFT torsional profiles (orange, green, and purple dashed curves) almost overlap with the reference profiles (black curves). 
As depicted in Figure~\ref{fgr:summary}, the DFT MAEs are extremely small (approximately 1~kJ/mol or less), which is consistent with previous studies.\cite{CKK97,KCK97,GMTKN55}
In the absence of molecular conjugation, such as in the CTB33-45 and CTB32-190 subsets, the torsional energy is primarily determined by steric hindrance between adjacent substituents,\cite{GS05,MG07,W20} indicating that steric hindrance does not cause significant DFT torsional errors.

\begin{figure}[h!]
\centering
\includegraphics[width=\columnwidth]{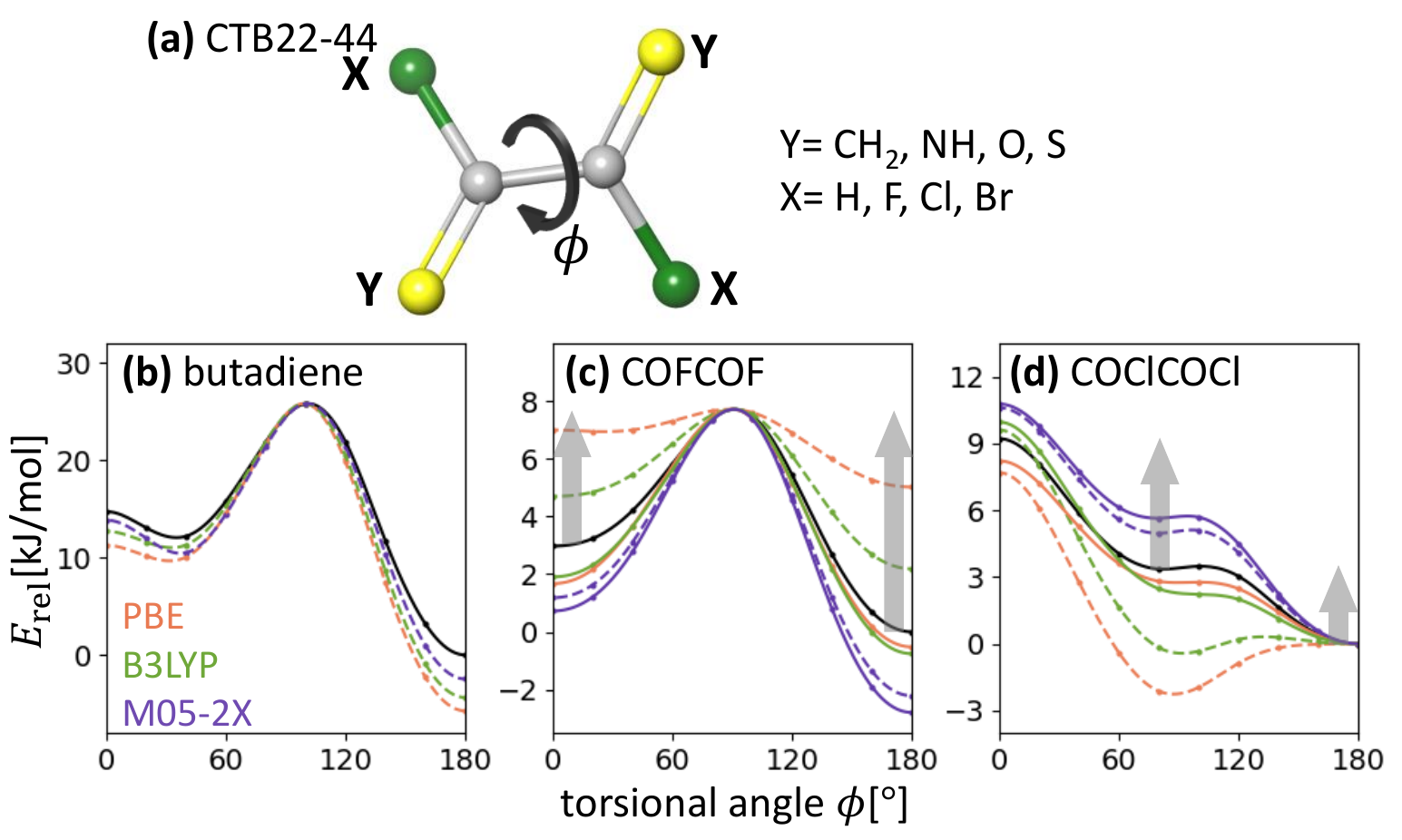}
\caption{
Failures of semilocal functionals for CTB22-44:
(a) Schematic.
Torsional angle $\phi$ is defined as the dihedral angle of Y-C-C-Y.
Torsional profile of (b) butadiene (CH$_2$CHCHCH$_2$), (c) oxalyl fluoride (COFCOF), and (d) oxalyl chloride (COClCOCl) obtained from the reference calculation (black), DFT methods (colored dashed lines), and HF-DFT methods (colored solid lines).
All energies are relative with zero for reference at global minimum, approximations match reference maximum in (b) and (c), and $\phi = 180^\circ$ in (d).
The gray arrows in (c) and (d) denote some barriers corresponding to CTB22ds-12.
}
\label{fgr:conj}
\end{figure}

\begin{figure}[h!]
\includegraphics[width=0.95\columnwidth]{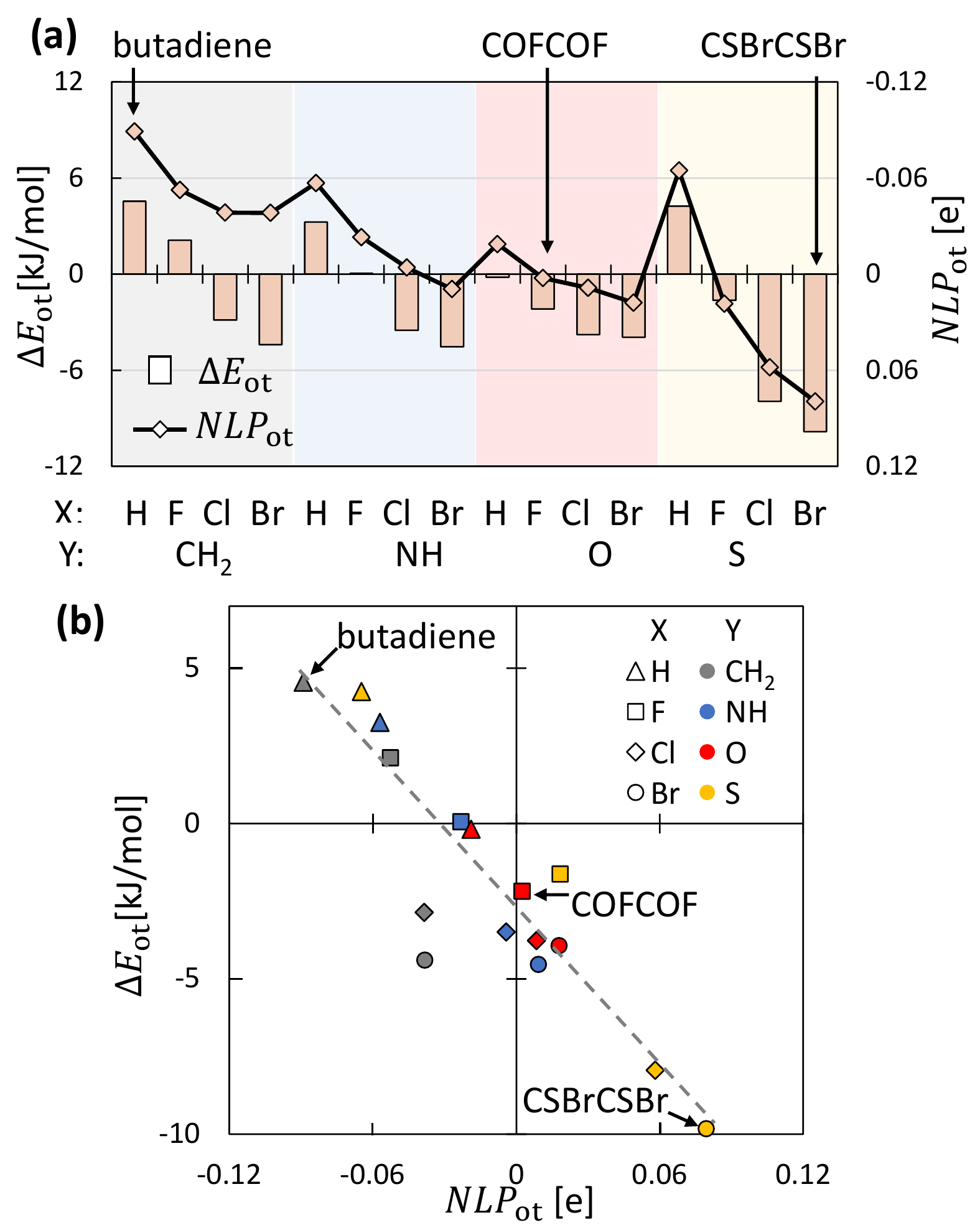}
\centering
\caption{
For B3LYP calculations of CTB22-44 molecules, 
(a) error of ortho-trans energy differences ($\Delta E\ot$, bars), and corresponding non-Lewis population ($NLP\ot$, a line with markers). 
Note that $NLP\ot$ is drawn in reverse for easy comparison with $\Delta E\ot$.
(b) Correlation between $NLP\ot$ and $\Delta E\ot$.
The gray dashed line is a guide for the eye.
All properties ($\chi$) are calculated as ortho-trans, $\chi\ot = \chi(ortho) - \chi(trans)$.
}
\label{fgr:ot}
\end{figure}

The torsional barrier of a single bond that participates in $\pi$-conjugation (i.e., $\22$), such as molecules in Figure~\ref{fgr:conj}a, is well known to be significantly more problematic in DFT than those without conjugation,\cite{CKK97,KCK97,SPM01,ZT06, SCB10, SKGB14, BW14,TBRL18} as seen in Figure~\ref{fgr:summary}.
The use of DFT on these types of torsional barriers needs caution.
In a $\pi$-conjugated molecule, the torsional barrier is determined by the competition between conjugation and steric hindrance.\cite{LL15}
When the molecules are planar, such as $cis$- ($\phi=0^\circ$) or $trans$ conformation ($\phi=180^\circ$), $\pi$-conjugation is formed and stabilizes the molecules. In contrast, no $\pi$-conjugation is formed for the $ortho$ conformation ($\phi\approx90^\circ$).

Butadiene is one of the simplest molecules that exhibits $\pi$-conjugation. Its torsional profiles are depicted in Figure~\ref{fgr:conj}b. 
Relative to the reference (black solid line), all DFT methods (colored dashed lines) in the figure indicate an overestimation of torsional barriers between the $ortho$ and $trans$ conformations.
This has been studied extensively and its physical origin is well known:
local or semilocal exchange hole models (commonly used in many DFT methods) underestimate the exchange energy density when the exchange hole is delocalized.\cite{B00,SPM01}
 This results in the overstabilization of the delocalized ($\pi$-conjugated) conformations.
In other words, a delocalized electronic structure is responsible for the delocalization error in DFT.

However, this is not true for all DFT torsional profiles.
Recently, Tahchieva $et~al.$ showed that glyoxal, oxalyl halides, and their thiocarbonyl derivatives have large DFT errors in their torsional profiles.
These errors are not due to the missing of dispersion interactions in DFT.\cite{TBRL18}
Moreover, although not explicitly stated, the direction of the DFT error of the molecules covered in that work is the opposite of typical DFT errors.
For example, for COFCOF (oxalyl fluoride) shown in Figure ~\ref {fgr:conj}c, the shape of the torsional profile is similar to that of butadiene; semi-local and hybrid functionals with moderate exact exchange predict the energy of the planar form to be relatively high. (See Figure~\ref{S-fgr:OF}a)
This is contrary to what would be expected in butadiene. Therefore, such a behavior cannot be explained by the delocalization error of the $\pi$-conjugated form.
COClCOCl (oxalyl chloride) suffers a more serious problem.
As depicted in Figure~\ref{fgr:conj}d, PBE and B3LYP incorrectly predict $ortho$ as the equilibrium conformation. (Figure~\ref{S-fgr:OCl}a and ~\ref{S-fgr:OBr}a confirms this also happens for semi-local and hybrid functionals with moderate exact exchange.)
Knowledge about the molecules that have the opposite tendency for DFT errors is lacking; in particular, it is unclear as to why DFT acts atypically on these molecules.

The CTB22-44 subset consists of 44 torsional barriers of 16 molecules, including 8 molecules discussed in the reference \citenum{TBRL18} which exhibit the opposite DFT error trend.
To simplify the analysis, we focus on the differences between the $ortho$ and $trans$ conformations ($\chi\ot = \chi(ortho) - \chi(trans)$ for a computable property $\chi$).
The $ortho$ and $trans$ conformations are expected to exhibit the most different physical and chemical properties, regardless of the detailed shape of the whole torsional profile.
Any method that yields accurate $E\ot$ (that is, the energy difference between the $ortho$ and $trans$ conformations, which becomes a torsional barrier when $ortho$ and $trans$ conformations are the local maximum and nearest minimum on the torsional profile, respectively) should show good performance on the whole torsional profile, including any barriers.

In Figure~\ref{fgr:ot}a, the B3LYP error trend of the $ortho$-$trans$ energy difference ($\Delta E\ot = E\ot^{\sss B3LYP} - E\ot^{\sss Ref}$) is consistent with that in Figure~\ref{fgr:conj}b$\sim$d; $\Delta E\ot > 0$ (B3LYP overestimates) for butadiene and $\Delta E\ot < 0$ (B3LYP underestimates) for oxalyl halides.
When the Y group changes as CH$_2 \rightarrow$ NH $\rightarrow$ O $\rightarrow$ S and X group changes as H $\rightarrow$ F $\rightarrow$ Cl $\rightarrow$ Br, $\Delta E\ot$ becomes increasingly negative, indicating the increasing overstabilization of the $ortho$ conformation (relative to $trans$) by B3LYP.
This is not limited to B3LYP. Other DFT methods follow the same trend (See Figure~\ref{S-fgr:Eot_all}).
The overall size of $\Delta E\ot$ in Figure~\ref{fgr:ot}a is small on an absolute scale, but is large enough to yield incorrect equilibrium conformation predictions in some cases, such as COClCOCl in Figure~\ref{fgr:conj}d.

To investigate the source of the DFT error, we performed natural bond orbital (NBO) analysis\cite{NBO31} on the wavefunctions of the molecules in CTB22-44.
In the NBO scheme, most of the electrons are strictly localized to the Lewis-type core, bonding, and lone-pair NBOs.
The remaining “non-Lewis” electrons, occupying antibonding or Rydberg NBOs, represent the delocalization correction to the idealized local Lewis structure,\cite{NBO3} so the degree of delocalization can be quantified using the non-Lewis electron population (NLP; the number of non-Lewis electrons).
In Figure~\ref{fgr:ot}a, the markers connected with a line indicate the NLP difference between $ortho$ and $trans$ conformations ($NLP\ot =NLP_{ortho} - NLP_{trans}$) obtained from the B3LYP calculations.
From this perspective, the sign of $NLP\ot$ indicates the relative degree of delocalization between $ortho$ and $trans$ conformations.
For example, when the $trans$ conformation is more delocalized than the $ortho$ conformation, $NLP_{trans} > NLP_{ortho}$ and thus $NLP\ot < 0$, and $vice~versa$. (Split sentences)
For the molecules in CTB22-44 subset, the almost linear relationship in Figure~\ref{fgr:ot}c indicates a strong correlation between the delocalization error ($\Delta E\ot$) and degree of electron delocalization ($NLP\ot$).

For butadiene (top left on Figure~\ref{fgr:ot}b), the negative sign of $NLP\ot$ matches the chemical intuition; the $\pi$-conjugated $trans$ conformation is more delocalized when compared with the $ortho$ conformation.
In Figure~\ref{fgr:ot}a, when the Y group changes as the CH$_2 \rightarrow$ NH $\rightarrow$ O $\rightarrow$ S and X group changes as H $\rightarrow$ F $\rightarrow$ Cl $\rightarrow$ Br, the $NLP\ot$ becomes increasingly positive, indicating that electrons in the $ortho$ conformation become increasingly delocalized than those in the $trans$ conformation.
However, for CSBrCSBr (bottom right), $NLP\ot > 0$ implies that the $ortho$ conformation exhibits greater electron delocalization than the $\pi$-conjugated $trans$ conformation.
This observation is somewhat unexpected and implies that both the $\pi$-conjugation in the planar form and the hyperconjugation maximized in a nonplanar form have a significant influence on the electron delocalization of the molecules in CTB22-44.
(We use the term hyperconjugation to denote any type of electron delocalization induced by the orbital interaction between the filled and antibonding orbitals.)
This observation also indicates that the electron delocalization due to the $\pi$-conjugation in the planar conformation may not always have a higher strength than the hyperconjugation in twisted conformations.
The sign of $NLP\ot$ in Figure~\ref{fgr:ot} is due to the competition between two delocalization factors.
When the delocalization in the $trans$ conformation ($\pi$ conjugation) prevails over the delocalization in the $ortho$ conformation (hyperconjugation), $NLP\ot < 0$, and $vice~versa$.

\begin{figure}[h!]
\includegraphics[width=0.9\columnwidth]{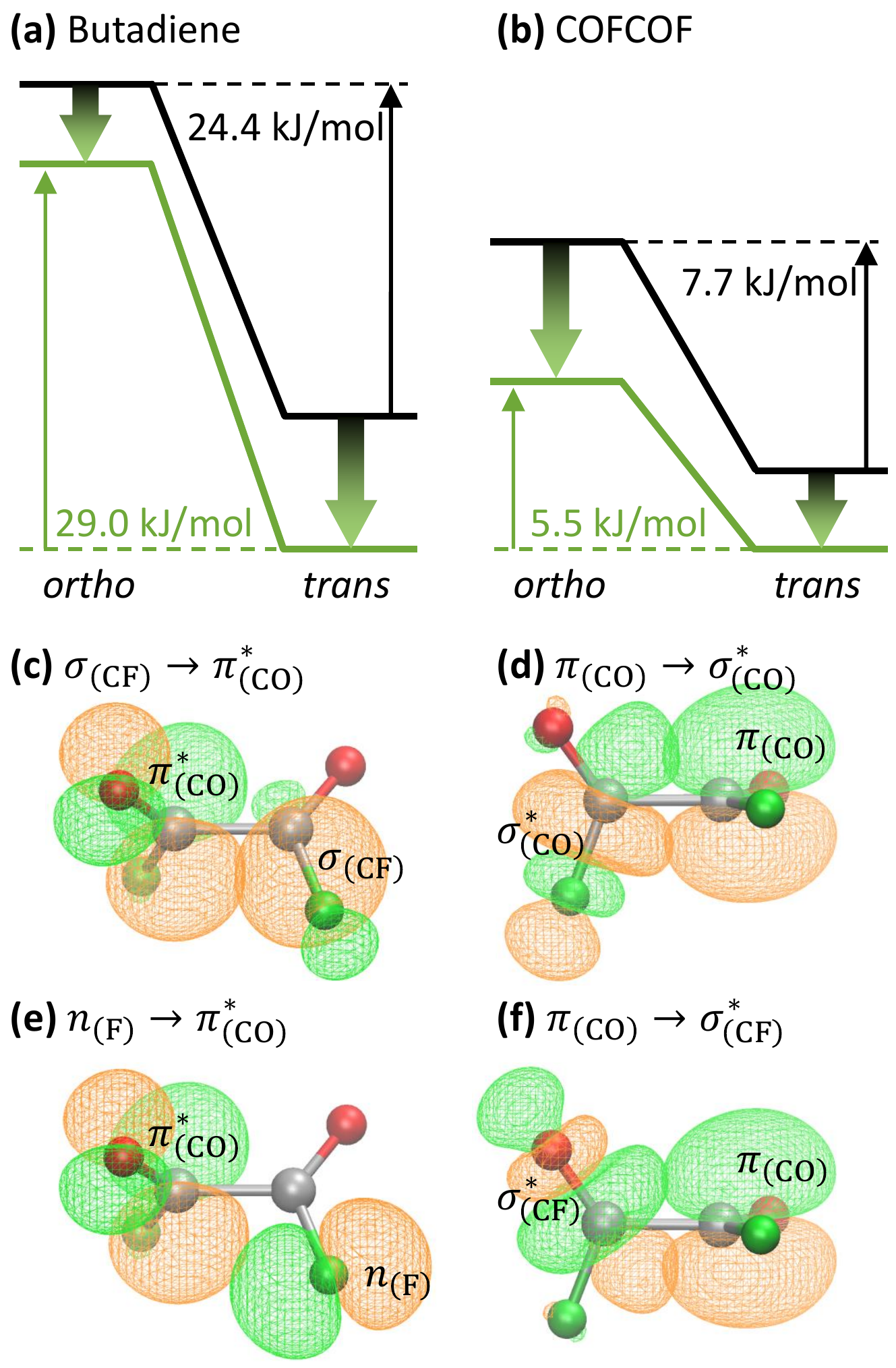}
\centering
\caption{
Schematics of $E\ot=E_{ortho}-E_{trans}$ for (a) butadiene and (b) oxalyl fluoride (COFCOF).
Black (green) energy levels denote the reference (B3LYP), whereas arrows with gradation indicate the delocalization error.
The size of arrows is not to scale.
(c$\sim$f) shows the selected NBO donor-acceptor pairs with maximum energy stabilization in the $ortho$ conformation of COFCOF. The isovalue is set to $\pm$0.05 electrons/bohr$^3$.
}
\label{fgr:nbo_scheme}
\end{figure}

Figures~\ref{fgr:nbo_scheme}a and \ref{fgr:nbo_scheme}b show schematics of $E\ot$.
A smaller delocalization error in the $ortho$ conformation than that in the $trans$ conformation of butadiene incorrectly increases $E\ot$ from 24.4~kJ/mol (reference) to 29~kJ/mol (B3LYP).
Conversely, in COFCOF, the delocalization error of the $ortho$ conformation is greater than that of the $trans$ form, reducing $E\ot$ from 7.7~kJ/mol (reference) to 5.5~kJ/mol (B3LYP).
Figures~\ref{fgr:nbo_scheme}c$\sim$f depict the representative hyperconjugative interactions maximized at $ortho$ conformation for the COFCOF molecule.
Similar types of hyperconjugation exist for other molecules; however, $n_{\sss (X)} \rightarrow \pi^*_{\sss (C=Y)}$ exists only for halide molecules.
For the CTB22-44 molecules, in which the X group is hydrogen, hyperconjugative interaction, such as those shown in Figure~\ref{fgr:nbo_scheme}e, does not exist and the $\pi$ conjugation is relatively more dominant than in their halogen counterparts.
The absence of this hyperconjugation for X=H results in negative $NLP\ot$ (i.e., the $ortho$ conformation is relatively less delocalized), as shown in Figure~\ref{fgr:ot}.
Some molecules in CTB22-44 are known to exhibit hyperconjugation,\cite{CK00,CK01}
and these types of hyperconjugations contribute to electron delocalization in the $ortho$ conformation, possibly increasing the delocalization error in the DFT calculations.
Therefore, the explanation that ‘‘a delocalized electronic structure is responsible for the large delocalization error of DFT,'' which holds for butadiene,\cite{CKK97,KCK97,SPM01} also holds for molecules in CTB22-44.
The difference in the DFT tendency arises from the difference in the more delocalized conformations.

As depicted in Figure~\ref{fgr:summary}, a simple solution, called HF-DFT, can improve the performance of conventional functional approximations (especially for the CTB22ds-12 subset) regardless of the source of the error.
The density sensitivity of a given DFT calculation is defined as~\cite{SSB18}
\begin{equation}
S=| \tilde{E}[n\LDA] - \tilde{E}[n\HF] |,
\label{eq:S}
\end{equation}
where $n\LDA$ and $n\HF$ are the density from the local density approximation and HF, respectively, and $\tilde{E}[\cdot]$ is an energy of some density functional approximation.
Depending on the property of interest, $\tilde{E}$ in Eq.~\ref{eq:S} can be the energy of a given conformation (e.g., conformational sensitivity, $S(\phi)$) or the energy difference between adjacent local maximum and minimum conformations (e.g., barrier sensitivity, $S\b$).
Equation~\ref{eq:S} measures the energy difference between two extreme nonempirical densities.
A small $S$ implies that the use of any reasonably accurate density to calculate the DFT energy will not significantly change its energetic error.
A large value of $S$ indicates density sensitivity and then often the use of HF-DFT improves the results.\cite{SSB18,KSSB18}
In CTB22ds-12, barrier sensitivities are greater than 1.5 kJ/mol in B3LYP (or 4~kJ/mol for PBE).
Sometimes, only a subset of a molecule's barriers are density sensitive.
Figure~\ref{fgr:summary} depicts that for the density-sensitive CTB22ds-12 subset (significantly more than other subsets, see triangles), HF-DFT significantly reduces MAE over conventional DFT.
Other subsets are density insensitive, and thus the reduction in error with HF-DFT is small.
In this sub-subset, HF-DFT versions of PBE, BP86, TPSS, TPSSH, and PBE0 all outperform any conventional DFT considered in this work. (See Figure~\ref{S-fgr:stats} and Table~\ref{S-table:MAE} for statistics.)
Figure~\ref{fgr:conj}d depicts that, for COClCOCl molecule, where barrier sensitivites for two major barriers (gray arrows in Figure~\ref{fgr:conj}d) exceed 1.5~kJ/mol with B3LYP calculations, HF-PBE and HF-B3LYP fix the minima in the torsional profile, as do other functionals (See Figure~\ref{S-fgr:OCl} and Figure~\ref{S-fgr:OBr} for COBrCOBr).
Indeed, HF-DFT shows comparable performance compared to the torsion-corrected atom-centered potential correction method introduced in the reference \citenum{TBRL18}. (See Table~\ref{S-table:Anatole} )

\begin{figure}[h!]
\includegraphics[width=0.7\columnwidth]{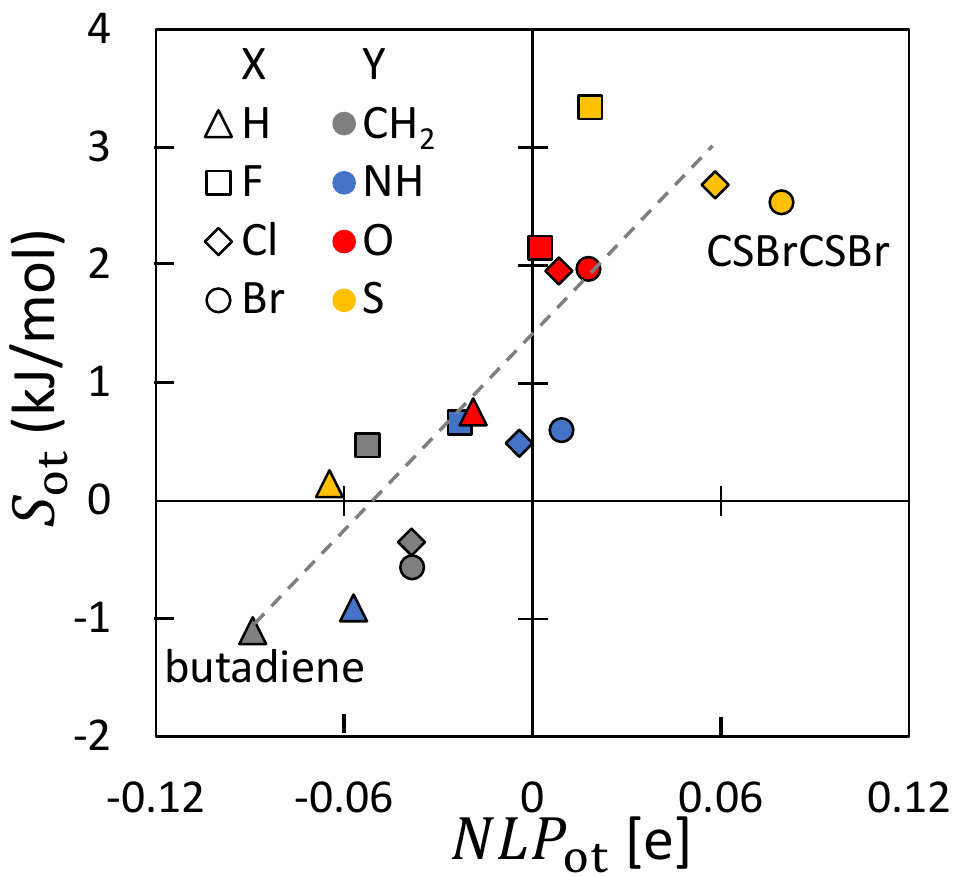}
\centering
\caption{
Correlation between $NLP\ot$ and $S\ot$ obtained from B3LYP calculations.
Note that $S\ot$ can take negative values.
The gray dashed line is a guide for the eye.
}
\label{fgr:NLP_S}
\end{figure}

Figure~\ref{fgr:NLP_S} depicts how the electron delocalization correlates with density sensitivity.
To compare relative amount of sensitivity of two conformations, we define $S\ot = S(ortho) - S(trans)$, the relative conformational sensitivity between $ortho$ and $trans$ conformations.
A positive $S\ot$ indicates that for the given molecule, $ortho$ conformational sensitivity is higher than that of $trans$ and $vice~versa$.
In most conformations, the signs of $NLP\ot$ and $S\ot$ are the same---for example, butadiene and CSBrCSBr, the two molecules at the extreme from the perspective of $NLP\ot$. 
More delocalized conformations are more density sensitive.

Although delocalization due to hyperconjugation is considerable on $\22$ molecules, it is also present in $\mix$ molecules.\cite{W20}
Hyperconjugation affects the DFT barrier error on CTB32-190, and the barrier error (although reduction is small) is observed to be reduced using HF-DFT, as shown in Figure~\ref{fgr:summary}.
We note that the procedure in HF-DFT is recommended to use HF densities only for density sensitive cases,\cite{KSSB18} but Figure~\ref{fgr:summary} shows using HF densities in all cases, showing this does not harm in density insensitive cases.

\begin{figure*}[h!]
\includegraphics[width=2\columnwidth]{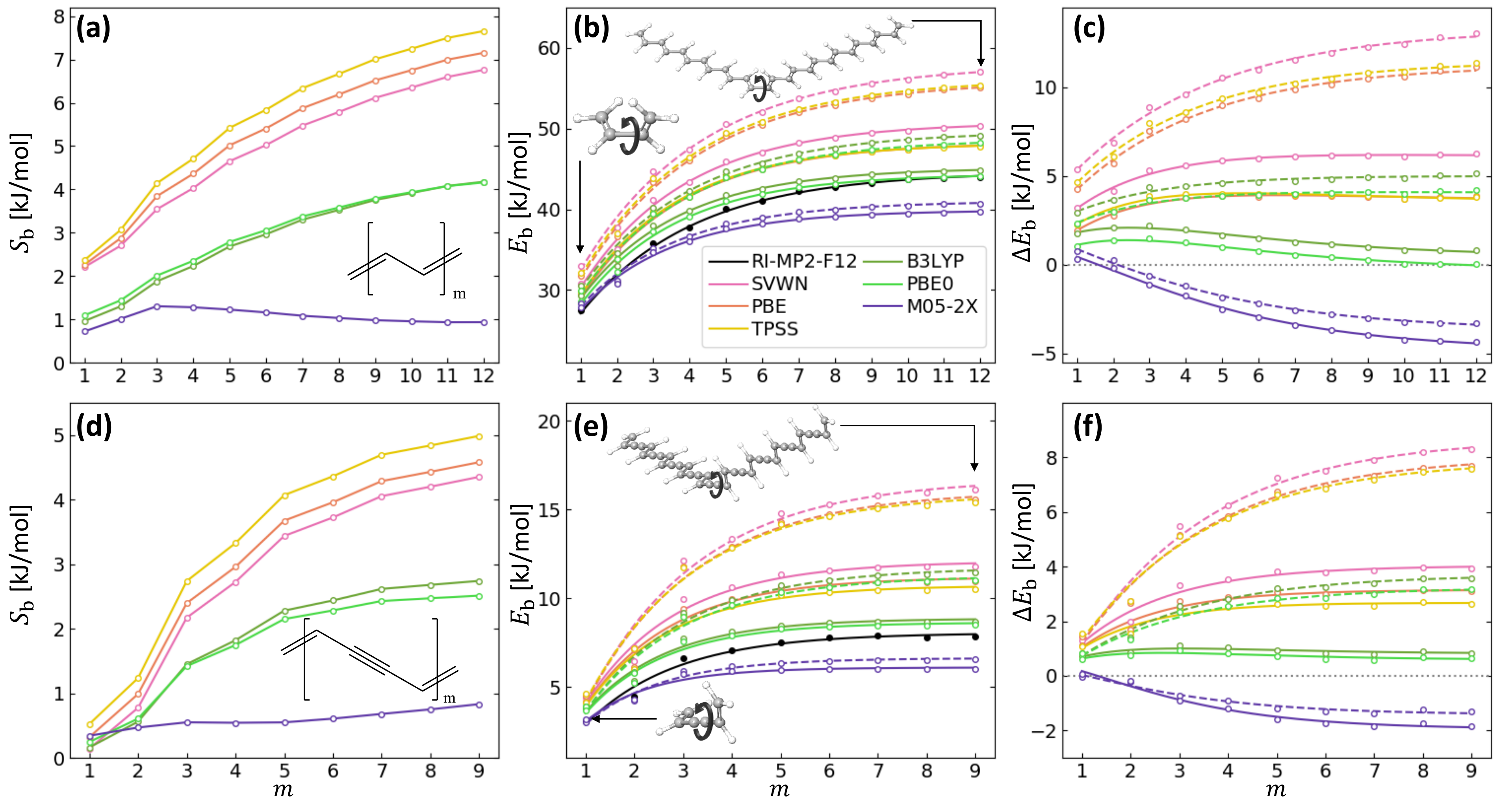}
\centering
\caption{
(a) Barrier sensitivity of polyacetylene (\ch{CH_2(C_2H_2)_{m}CH_2}), where $m$ is the number of repeated units. 
(b) Comparison of DFT (dashed lines) and HF-DFT (solid line) barriers with reference (black line) RI-MP2-F12 calculation.
(c) $E\b$ error relative to the reference. 
(d$\sim$ f) The same data but for polydiacetylene (\ch{CH_2 (CHC+CCH)_m CH_2}).
Following reference \citenum{SKGB14}, barriers are defined as $E\b = E(100^\circ)-E(180^\circ)$ for polyacetylene, and $E\b = E(90^\circ)-E(180^\circ)$ for polydiacetylene.
The curves in (b, c, e, f) are generated using the effective conjugation length model fitting in the reference \citenum{MSK97}.
}
\label{fgr:polymer}
\end{figure*}

Our analysis is not limited to small molecules.
Figure~\ref{fgr:polymer}a depicts the barrier sensitivity of polyacetylene (\ch{CH_2(C_2H_2)_{m}CH_2}) for various lengths ($m$) and various DFT methods.
For most functionals, $S\b$ increases with the increase in polymer length $m$.
This corresponds to extending the gray solid line in Figure~\ref{fgr:NLP_S} in the lower-left direction. (Note that $S\b$ is always positive while $S\ot$ is not.)
In other words, a longer polyacetylene exhibits larger electron delocalization of the $\pi$-conjugation type, and this increases $S\b$ for most functionals.
An exception is M05-2X because the functional has a significantly high portion of exact exchange.

Torsional barriers of polyacetylene molecules computed using DFT and HF-DFT and their errors are depicted in Figures~\ref{fgr:polymer}b and ~\ref{fgr:polymer}c, respectively.
DFT torsional barriers are typically overestimated for the same reasons as for butadiene, and the overestimation is larger for longer polyacetylenes owing to the stronger $\pi$ conjugation (i.e., delocalization error).
Again, for torsional barriers with a significant barrier sensitivity, HF-DFT effectively reduces the error of DFT.
The increase in error with the increase in $m$ is smaller for HF-DFT than for conventional DFT, and HF-B3LYP outperforms M05-2X for long polyacetylene torsional barriers.
This trend also applies to polymers containing triple bonds of $sp$ hybridization.
The lower panel in Figure~\ref{fgr:polymer} depicts the torsional barriers of various polydiacetylene (\ch{CH_2 (CHC+CCH)_m CH_2}) molecules.
Although polydiacetylenes have different types of hybridization, $sp^2$-$sp$-$sp$-$sp^2$, the physical origin of barrier overestimation by DFT is the same as that for polyacetylenes or butadiene.
Figures~\ref{fgr:polymer}c and f show that for these polymers, the performance of 100\% long-range  exact-exchange functionals is not optimal, as is known for aromatic compounds.\cite{K17}
Also, the excellent performance for short oligomers with M05-2X  disappear as $m$ increases, while the error of HF-PBE0 and HF-B3LYP saturates or even decreases.
The HF-DFT version of hybrid functionals with a moderate amount of exact exchange (B3LYP, PBE0, and M06, see Figure~\ref{S-fgr:polymer_supp}) shows the best performance in the torsional barrier of the polymer considered in this work.

To conclude, in this work, we investigated the carbon--carbon torsional profiles of various molecule types using standard DFT.
In general, the DFT error in the torsional barrier is larger for $\pi$-conjugated molecules than for nonconjugated molecules.
Typically, for $\pi$-conjugated molecules, planar conformations are overstabilized by DFT and so torsional barriers are overestimated.
But there exist some atypical torsional profiles where non-conjugated conformations are overstabilized by DFT, causing significant barrier errors.
We demonstrate that this atypical error behavior originates from electron delocalization due to strong hyperconjugation.
The error in DFT torsional profiles of conjugated molecules is determined by the competition of two electron-delocalization factors---$\pi$-conjugation and hyperconjugation---whose relative strength varies with conformation, and becomes severe when one factor dominates the other.
For oxalyl or thiocarbonyl halides, hyperconjugation dominates, while for long conjugated molecules, $\pi$-conjugation dominates, and both suffer from severe DFT errors.
All these poor actors are density sensitive, and so HF-DFT significantly improves the torsional profiles regardless of the type of electron delocalization.
This is also true for long polymers.
For density sensitive torsional barriers, applying HF-DFT on semi-local functionals even outperforms exchange-enhanced or range-separated hybrid functionals.
Taken all together, HF-PBE0 provides high accuracy for all small molecule torsional barriers, and its errors remain small for long-chain polymers.


\section*{Computational Details}
All ab-initio calculations were performed using the ORCA 4.2.1 program package.\cite{ORCA20}
Most of the reference calculations on the CTB-279 dataset were obtained at the DLPNO-CCSD(T)\cite{DLPNO16}/TightPNO/cc-pV(T+Q)Z level, and we confirmed that the use of the NormalPNO setting gives no meaningful difference.
Some exceptions were observed for X=F, Cl, and Br for the CTB22-44 subset.
Therefore, we used CCSD(T)/cc-pV(T+Q)Z for those molecules.
We used the VeryTightSCF keyword for all CC calculations to maintain high integral accuracy.
We confirmed that T1 diagnostics of all molecules in the CTB-279 dataset are all less than 0.02.
For polyacetylene and polydiacetylene, RI-MP2-F12/aug-cc-pVDZ was used.
All DFT and HF-DFT calculations (SVWN,\cite{D30,VWN80} PBE,\cite{PBE96} BP86,\cite{B88,P86} BLYP,\cite{B88,LYP88} TPSS,\cite{TPSS} TPSSH,\cite{TPSSH} B3LYP,\cite{B3LYP} PBE0,\cite{PBE0} M06,\cite{M06} BHHLYP,\cite{BHHLYP} M06-2X,\cite{M06} M05-2X,\cite{M052X} CAM-B3LYP,\cite{CAMB3LYP} $\omega$B97, $\omega$B97X,\cite{wB97} and $\omega$B97M-V\cite{wB97MV}) were performed with the PySCF program package\cite{PySCF18}.
For the CTB-279 dataset and polymer DFT calculations, Dunning's\cite{D89, D92, D93} cc-pVQZ and aug-cc-pVDZ basis sets were used, respectively.
Reference \citenum{HF-DFT} introduces simple scripts to run HF-DFT calculations.
The NBO analysis was performed using the NBO 3.1 program\cite{NBO31} included in Gaussian 16.\cite{G16}

To generate the CTB-279 dataset, molecules were generated automatically by attaching various functional groups (R, X, and Y in Figure~\ref{fgr:noconj} and \ref{fgr:conj}).
In this step, the initial conformation of each molecule was selected such that no intramolecular hydrogen bond was formed.
For example, two OH groups in glyoxal were forced to move in opposite directions.
Since MP2 yields very accurate geometries for covalently bonded systems\cite{VB20} and RI-MP2 gives very close geometry to MP2,\cite{RIMP2} we optimized the molecules using RI-MP2/def2-TZVP\cite{def2} with a fixed $\phi$ with $20^\circ$ intervals.
After optimization, we manually removed the molecules that formed intramolecular hydrogen bonds at any $\phi$.
In this process, 15, 66, and 16 molecules remained for the CTB33, CTB32, and CTB22 subsets, respectively.
With these molecules, we performed DLPNO-CCSD(T) or CCSD(T)/cc-pV(T+Q)Z calculations to obtain the reference energies.
With the reference torsional profile, we used cubic spline interpolation to calculate the local maximum and minimum in $1^\circ$ precision.
Torsional profiles in Figure~\ref{fgr:noconj} and \ref{fgr:conj} are also obtained with this procedure.
After locating the local maximum and minimum, geometry optimization with a fixed torsional angle was performed. On that geometry, reference and DFT calculations were performed.
As such, CTB33 and CTB32 contained 45 and 190 torsional barriers, respectively.


\vspace{5mm}
\begin{acknowledgement}
This study was performed at Yonsei University and was supported
by a grant from the Korean Research Foundation
(2020R1A2C2007468 and 2020R1A4A1017737). KB acknowledges the NSF for Grant CHE 1856165.\\
\end{acknowledgement}

\section*{Supporting Information}

$\bullet$ Figures and tables corresponding to ortho-trans energy differences, statistics for CTB-279 dataset, comparison with PBE-TCACP and HF-DFT, problematic torsional profiles, and polymer torsional barrier errors for various functionals. \\
$\bullet$ Spreadsheet containing CTB-279 and polymer data\\
$\bullet$ Geometry files used in this work \\

\bibliographystyle{ieeetr}
\bibliography{torsL_submit.bbl}


\clearpage
\clearpage

\beginsupplement
\section*{Supporting Information}

\begin{figure}[h!]
\includegraphics[width=1.0\columnwidth]{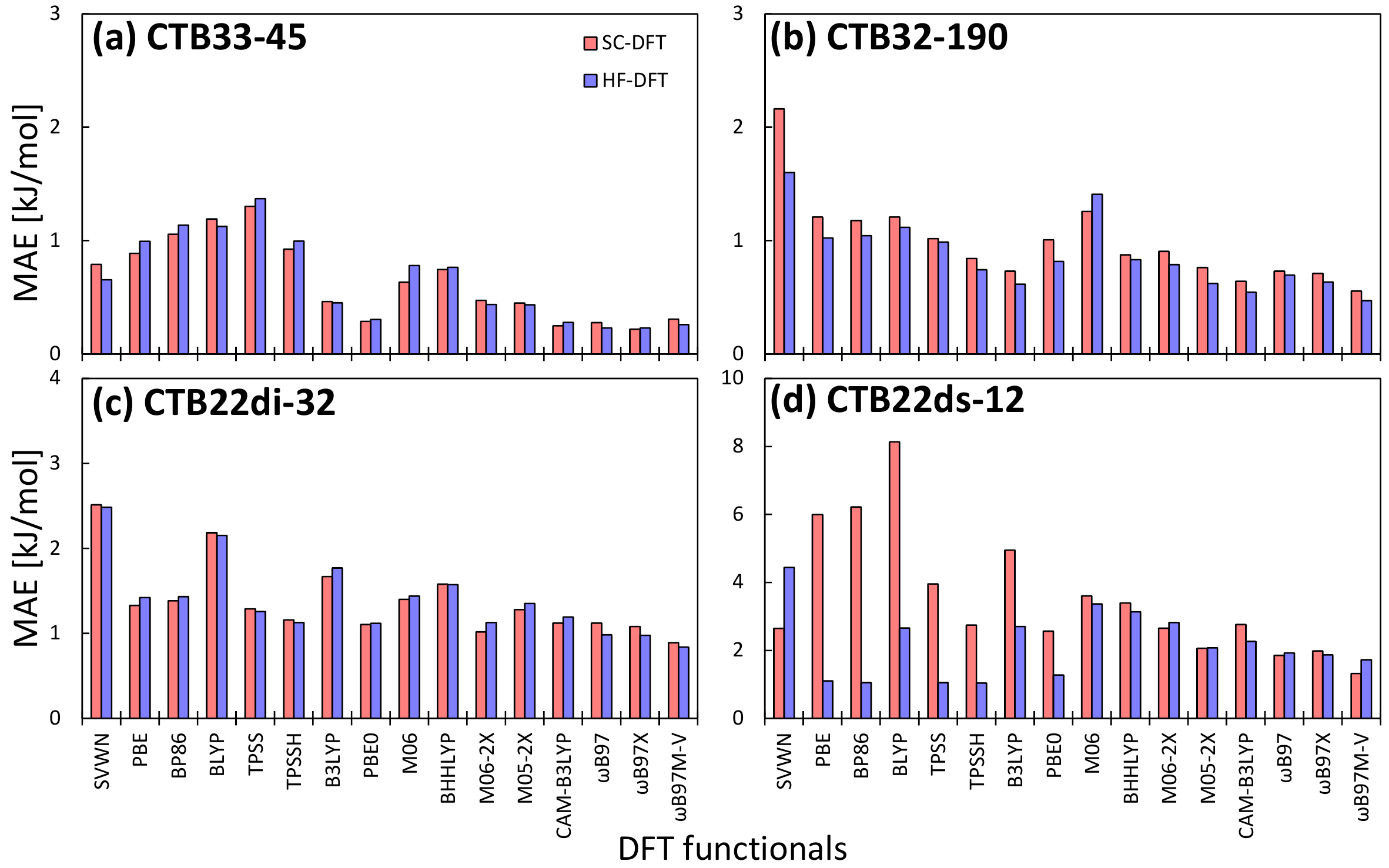}
\centering
\caption{
Torsional barrier MAEs of various functionals. 
Torsional barriers are defined as the energy difference between the local minimum and nearest local maximum in each torsional profile.
}
\label{fgr:stats}
\end{figure}

Table~\ref{table:MAE} and Figure~\ref{fgr:stats} compares DFT and HF-DFT for torsional barriers.
For highly density sensitive CTB22ds-12, HF-DFT shows dramatic improvement over conventional DFT, while for other density insensitive subsets, HF-DFT yields similar results with conventional DFT.

\begin{table}[h!]
\includegraphics[width=0.8\columnwidth]{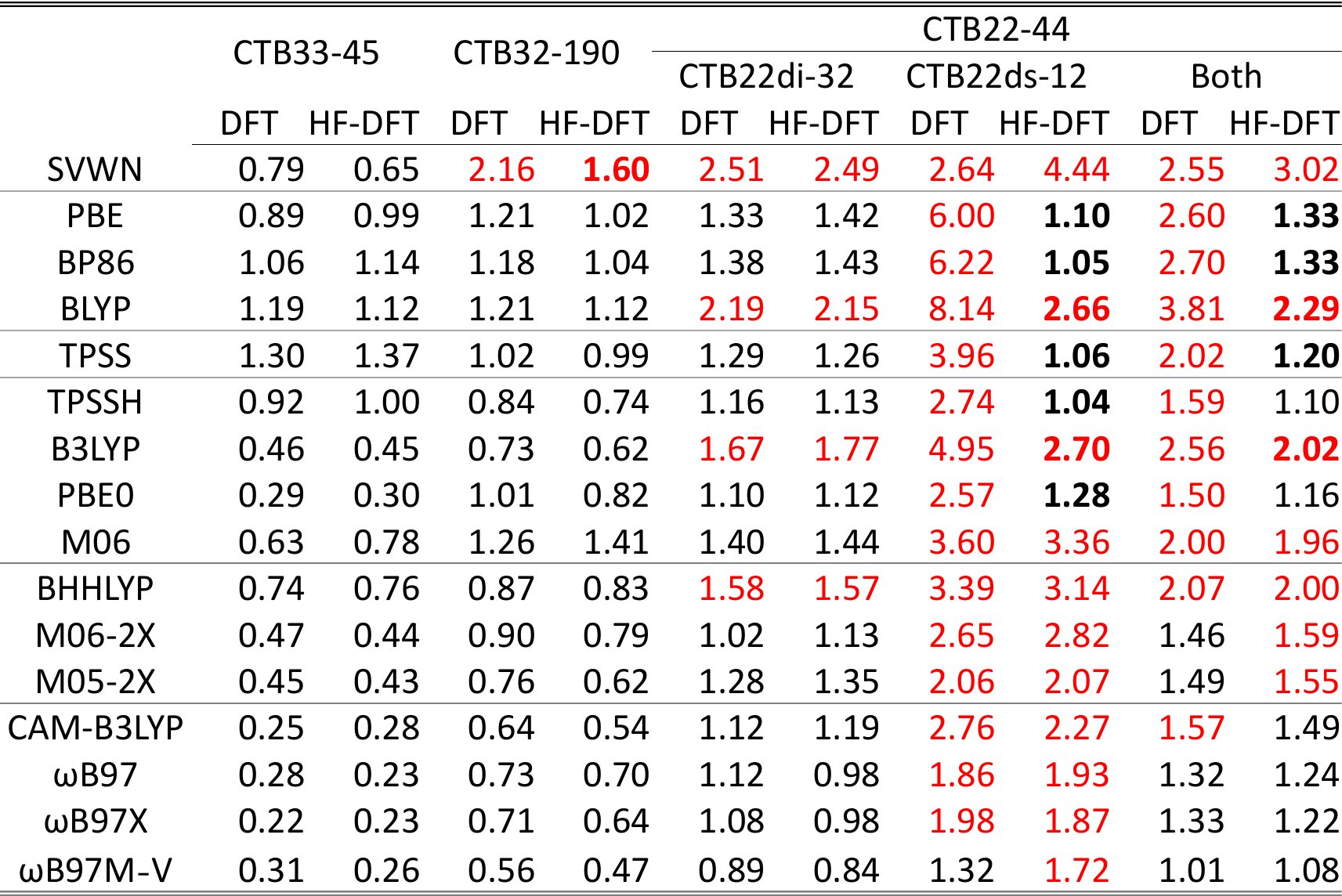}
\centering
\caption{
Torsional barrier MAEs (in kJ/mol) of various functionals.
Red numbers indicate MAE over 1.5~kJ/mol. 
Bold numbers indicate that HF-DFT reduces MAE by at least 0.5~kJ/mol compared to DFT.
}
\label{table:MAE}
\end{table}

\begin{figure}[h!]
\includegraphics[width=1.0\columnwidth]{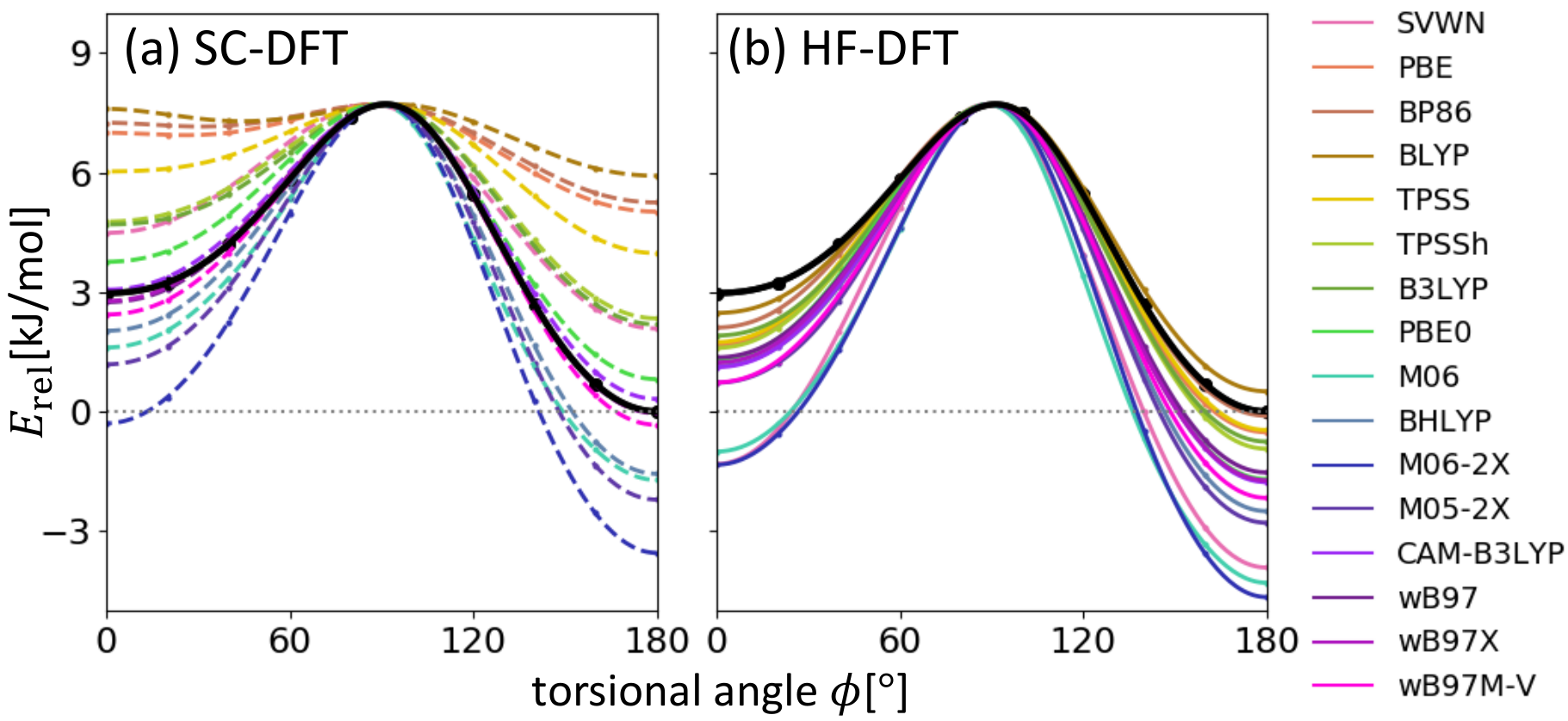}
\centering
\caption{
COFCOF (oxalyl fluoride) torsional profile for reference (CCSD(T)/cc-pV(T+Q)Z, black) and various (a) SC-DFT methods and (b) HF-DFT methods.
}
\label{fgr:OF}
\end{figure}

\begin{figure}[h!]
\includegraphics[width=1.0\columnwidth]{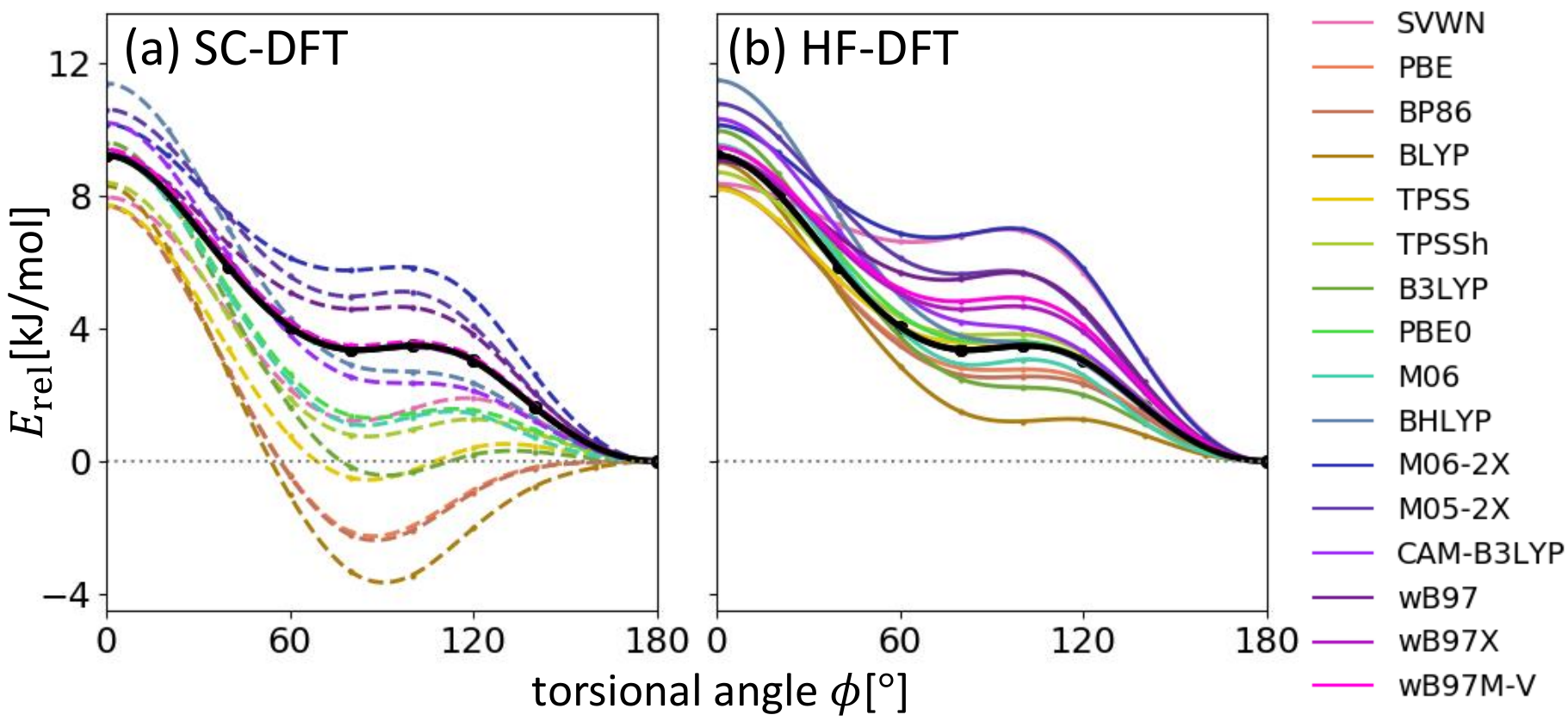}
\centering
\caption{
COClCOCl (oxalyl chloride) torsional profile for reference (CCSD(T)/cc-pV(T+Q)Z, black) and various (a) SC-DFT methods and (b) HF-DFT methods.
}
\label{fgr:OCl}
\end{figure}

\begin{figure}[h!]
\includegraphics[width=1.0\columnwidth]{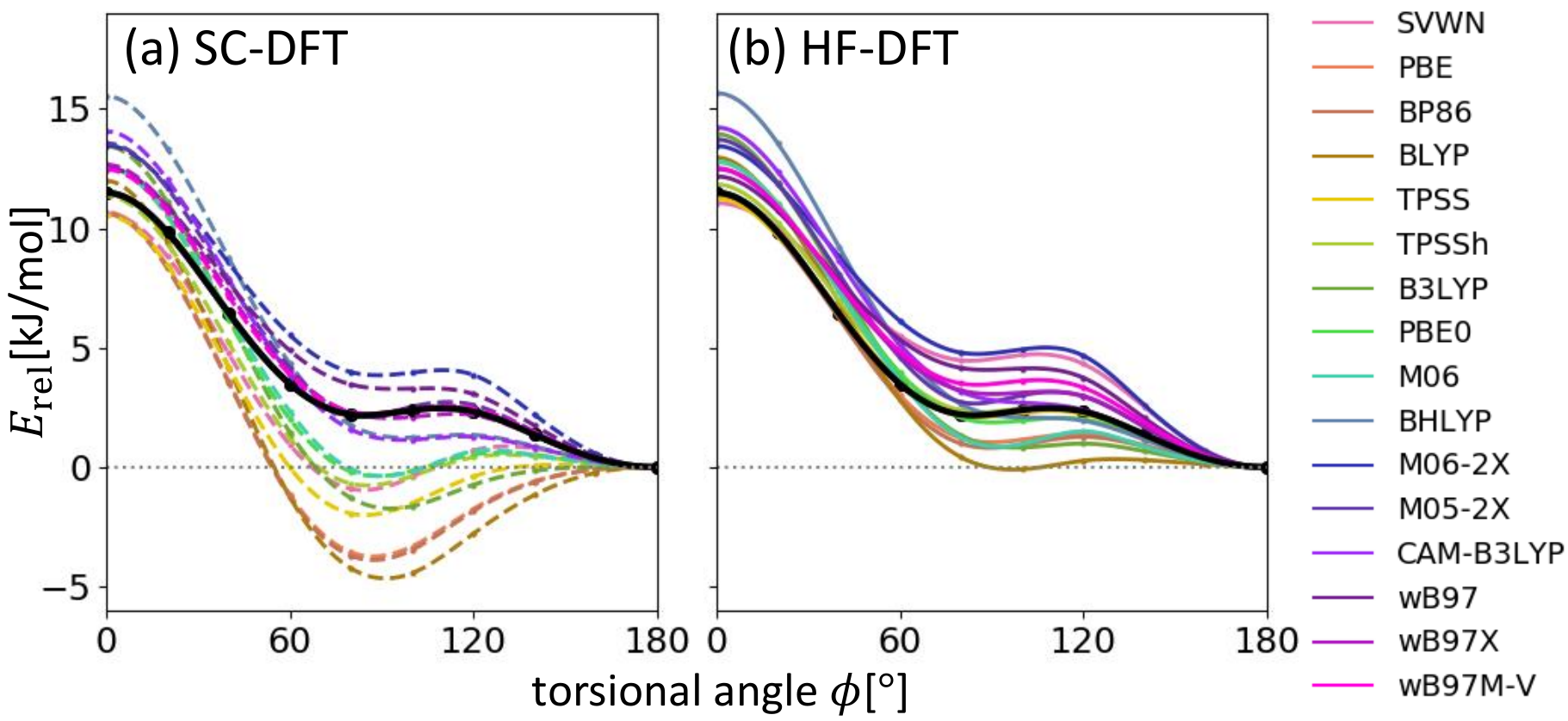}
\centering
\caption{
COBrCOBr (oxalyl bromide) torsional profile for reference (CCSD(T)/cc-pV(T+Q)Z, black) and various (a) SC-DFT methods and (b) HF-DFT methods.
}
\label{fgr:OBr}
\end{figure}

Figure~\ref{fgr:OF}, \ref{fgr:OCl}, and \ref{fgr:OBr} shows torsional profiles of COFCOF, COClCOCl, and COBrCOBr, respectively, obtained with conventional DFT and HF-DFT. 
Overall, $ortho$ conformations of these molecules are typically overstabilized with local/semi-local/hybrid functionals with moderate amout exchange.
While some conventional DFT predicts incorrect global minimum conformations for COClCOCl and COBrCOBr, HF-DFT captures correct global minimums.

\begin{figure}[h!]
\includegraphics[width=0.8\columnwidth]{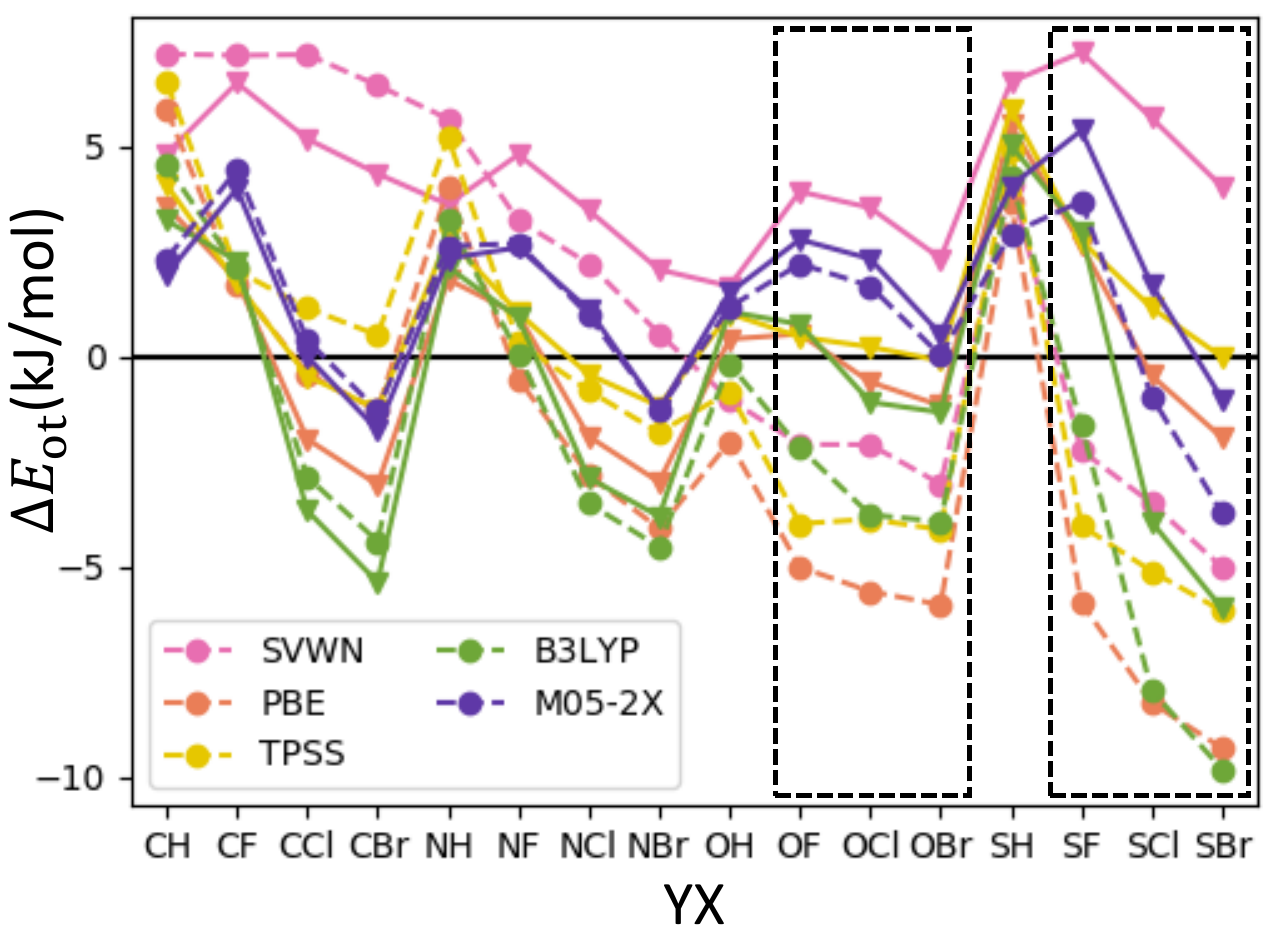}
\centering
\caption{
$\Delta E\ot$ for CTB22-44 molecules.
Squares with dashed lines denotes molecules containing CTB22ds-12 barriers.
Dashed lines with circle markers denotes DFT, while solid lines with triangle markers denotes HF-DFT.
}
\label{fgr:Eot_all}
\end{figure}

Figure~\ref{fgr:Eot_all} shows that various DFT methods, not only B3LYP, typically shows increasingly negative $\Delta E\ot$ when the Y group changes as CH$_2 \rightarrow$ NH $\rightarrow$ O $\rightarrow$ S and X group changes as H $\rightarrow$ F $\rightarrow$ Cl $\rightarrow$ Br.

Table~\ref{table:Anatole} compares performances between DFT, HF-DFT and PBE-TCACP\cite{TBRL18} on 12 conformational energy differences of some CTB22-44 molecules. 
HF-DFT with standard functionals shows comparable performances with empirically parameterized PBE-TCACP.

\begin{table}[h!]
\centering
\begin{tabular}{crr}
\hline \hline
Method     & \multicolumn{1}{c}{DFT} & \multicolumn{1}{c}{HF-DFT} \\
\hline
PBE-TCACP & \multicolumn{2}{c}{1.3}                              \\
SVWN       & 2.5                     & 3.6                        \\
PBE        & 5.0                     & 1.2                        \\
BP86       & 5.1                     & 1.2                        \\
BLYP       & 6.6                     & 2.7                        \\
TPSS       & 3.5                     & 1.2                        \\
TPSSH      & 2.5                     & 1.1                        \\
B3LYP      & 4.3                     & 2.7                        \\
PBE0       & 2.3                     & 1.2                        \\
M06        & 3.2                     & 2.7                        \\
BHHLYP     & 3.5                     & 3.2                        \\
M06-2X     & 2.3                     & 2.2                        \\
M05-2X     & 2.1                     & 2.1                       \\
CAM-B3LYP &2.5 & 2.3 \\
$\omega$B97 & 2.2 & 2.0 \\
$\omega$B97X & 2.2 & 1.9 \\
$\omega$B97M-V & 1.6 & 1.5 \\
\hline \hline
\end{tabular}
\caption{
Comparision of DFT, HF-DFT and PBE-TCACP (taken from \citenum{TBRL18}) results.
Shown numbers are MAEs of 12 energy differences between two torsional conformations (defined in \citenum{TBRL18}) of 8 molecules (Y=O,S with X=H,F,C,Br) used in both \citenum{TBRL18} and this work.
Reference to determine errors are taken from our calculations: DLPNO-CCSD(T)/TightPNO/cc-pV(T+Q)Z for X=H and CCSD(T)/cc-pV(T+Q)Z for others on RI-MP2/def2-TZVP geometries.
}
\label{table:Anatole}
\end{table}

\begin{figure*}[h!]
\includegraphics[width=1.8\columnwidth]{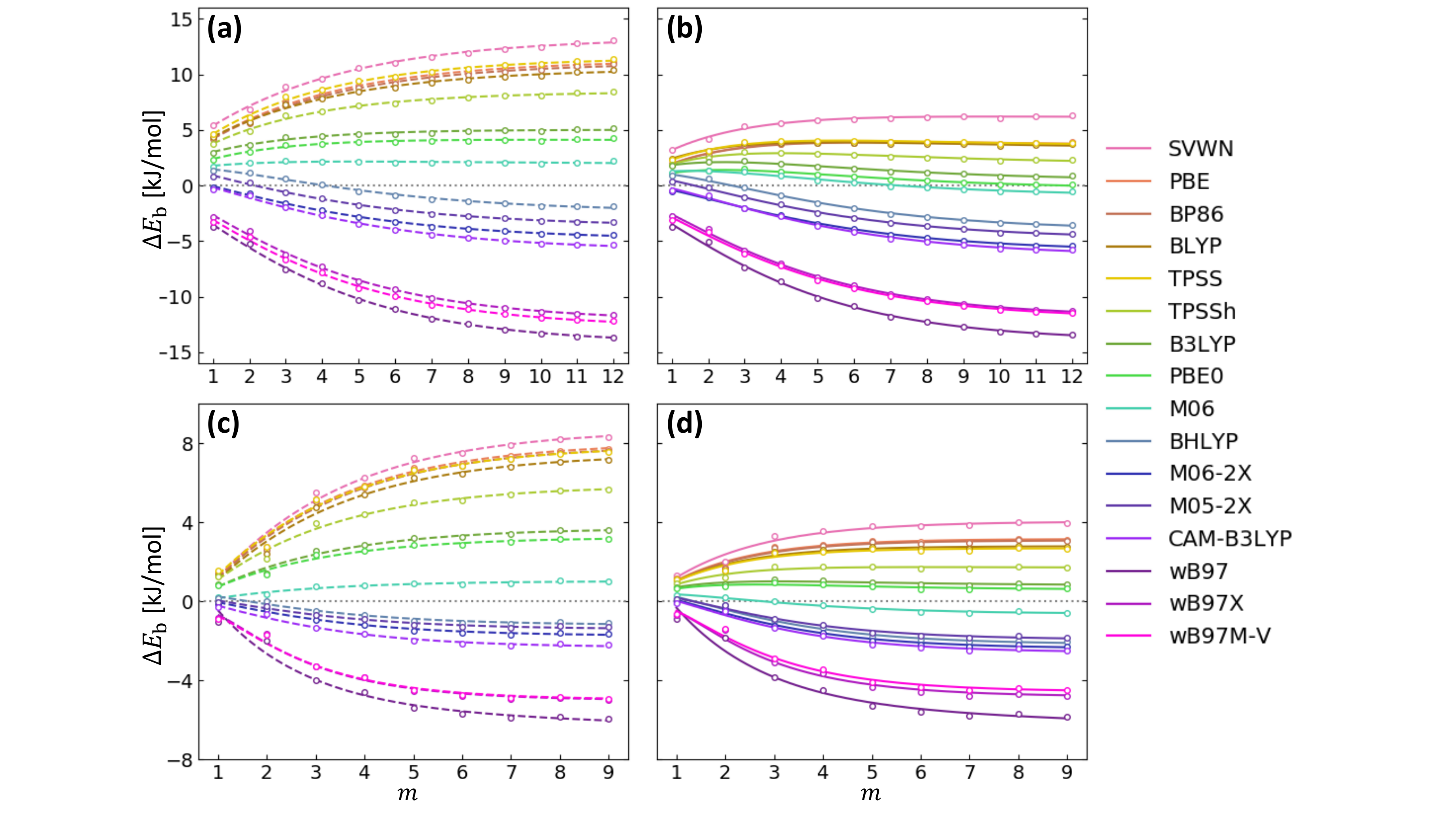}
\centering
\caption{
DFT and HF-DFT (left and right panel, respectively) $\Delta E\b$ for polyacetylene and polydiacetylene (upper and lower panel, respectively)
}
\label{fgr:polymer_supp}
\end{figure*}

Figure~\ref{fgr:polymer_supp} shows DFT and HF-DFT (left and right panel, respectively) $\Delta E\b$ for polyacetylene and polydiacetylene (upper and lower panel, respectively).
For short polymers, exchange-enhance functionals such as BHHLYP, M06-2X, and M05-2X show very small errors, but as $m$ increases, their error increases.

\end{document}